\documentclass[aastex]{emulateapj}
\tighten
\def\spose#1{\hbox to 0pt{#1\hss}}

\def\msun{{\rm ~M}_{\odot}}

\def\mpy{{\rm ~M}_{\odot} {\rm ~yr}^{-1}}
\def\lta{\mathrel{\spose{\lower 3pt\hbox{$\mathchar"218$}}
     \raise 2.0pt\hbox{$\mathchar"13C$}}}
\def\gta{\mathrel{\spose{\lower 3pt\hbox{$\mathchar"218$}}
     \raise 2.0pt\hbox{$\mathchar"13E$}}}

\def\be{\begin{equation}}
\def\ee{\end{equation}}

\begin{document}

\title{Magnetic Braking Revisited}

\author{N.~Ivanova \& Ronald E.~Taam}

\affil{  Northwestern University, Dept. of Physics \& Astronomy,
       2145 Sheridan Rd., Evanston, IL 60208\\
nata@northwestern.edu,  r-taam@northwestern.edu}

\begin{abstract}{

We present a description for the angular momentum loss rate due to magnetic braking 
for late type stars taking into account recent observational data on the relationship 
between stellar activity and rotation. The analysis is based on an idealized two 
component coronal model subject to constraints imposed on the variation of the 
coronal gas density with rotation period inferred from the observed variation of
X-ray luminosity, $L_{\rm x}$, with rotation rate, $\Omega$, ($L_{\rm x} \propto
\Omega^2$) for single rotating dwarfs.  An application of the 
model to high rotation rates leads to a gradual turnover of the X-ray luminosity  
which is similar to the saturation recently observed in rapidly rotating dwarfs. 
The resulting angular momentum loss rate, $\dot J$, depends on 
$\Omega$ in the form $\dot J \propto \Omega^{\beta}$ where $\beta \sim 3$ for slow 
rotators and $\sim 1.3$ for fast rotators.  The relation at high rotation rates significantly 
differs from the power law exponent for slowly rotating stars, depressing the angular 
momentum loss rate without necessarily requiring the saturation of the magnetic field.
The application of this model to the evolution of cataclysmic variable binary systems
leads to mass transfer rates that are in approximate accord with those observed in 
comparison to rates based on either a Skumanich law or an empirical law based on  
$\beta = 1$.}

\end{abstract}


\section{INTRODUCTION} 

It is generally accepted that the angular momentum removed from a stellar surface by the 
action of a magnetically coupled stellar wind is important in studies of the formation and 
%evolution of close binary systems.  Magnetic braking (MB) is the fundamental mechanism  
%ron's changes
evolution of close binary systems.  Magnetic braking (MB) is thought to be the 
fundamental mechanism responsible 
for orbital angular momentum loss in a number of classes of close binaries, such as  
cataclysmic variables (CVs) and low mass X-ray binaries (LMXBs). For such systems,
MB has two important roles. It provides a mechanism for promoting mass transfer from the 
low mass donor to its more massive compact companion.  In addition, it determines the 
range of post common envelope orbital separations for which the system evolves into 
this mass transfer stage. 

MB was first suggested as a mechanism for removing angular momentum from single stars 
by Schatzman (1962), who noticed that slowly rotating stars have convective envelopes.
A key insight was his recognition that material lost from the stellar surface is kept 
in corotation with the star by the magnetic field, leading to the result that the 
specific angular momentum carried by the gas is significantly greater than in a 
spherically symmetric stellar wind. To maintain  corotation, a star should possess
a substantial magnetic field. This can be achieved in low-mass main sequence (MS)
stars by gas motions in the deep convective envelope, leading to the generation
of a magnetic field by dynamo action.

Evidence for angular momentum loss in single stars is provided by the observations of 
young low-mass T Tauri stars which have shown that the stellar angular momentum 
is significantly higher in the pre-MS stage  than in the MS stage for  stars of 
similar masses (Stauffer \& Hartmann 1987).  In addition, the observations  of 
rotational velocities of stars in young clusters show a high rate of angular momentum
loss.  Specifically, the K-dwarfs in the Hyades (age $\approx 6 \times10^8$ yr) have 
mean rotational velocities below 10 km s$^{-1}$, but stars of the same spectral type 
in the Pleiades (age $\approx 7\times10^7$ yr) have mean rotational velocities of
about 40 km s$^{-1}$ (Stauffer 1987).  These observations supplemented the early work of
Skumanich (1972) who showed that the equatorial rotation velocities of G type MS stars
decrease with time, $t$, as $t^{-0.5}$. We note that this empirical dependence, upon 
which many studies are based, was established for stars with velocities
in the range of 1 to 30 km s$^{-1}$, but its applicability to the MS-like 
companion stars in close binaries of short orbital period (where rotational velocities 
are $\gta 100$ km s$^{-1}$) is suspect.

In parallel to these observational developments, many theoretical efforts were 
undertaken to understand the angular momentum loss mechanism
over a wide range in rotation rates. For example, the Skumanich relation for slowly 
rotating stars can be reproduced by an angular momentum loss rate, $\dot J$, 
proportional to $\Omega^3$, where $\Omega$ is the stellar rotational angular velocity, 
for a thermally driven stellar wind.  In this case, the magnetic field is assumed radial 
and to vary linearly with $\Omega$ (Weber \& Davis 1967). For fast rotation, 
a wide variety of theoretically predicted laws in the form $\Omega \propto t^{-\alpha}$ 
are possible with $\alpha$ ranging from 0.5 to 4 (Mestel 1984). More recently, Keppens,
MacGregor, \& Charbonneau (1995) considered a polytropic magnetized solar wind 
within the framework of a model developed by Weber \& Davis (1967) showing that the 
angular momentum loss rate is consistent with the Skumanich law at low angular velocities, 
but the power law dependence is shallower asymptotically approaching $\dot J \propto 
\Omega^2$ at high angular
velocities due to centrifugal effects. Additional factors affecting the angular
momentum loss have been  pointed out by Mestel \& Spruit (1987, hereafter MS87) who 
showed that, in a two-component model, more gas is confined in a dead zone, which
does not contribute to the braking, at high angular velocities. 

To further complicate the picture, significant observational developments   
have increased the disparity between observations and interpretations based on the 
Weber-Davis type prescription (for a review see Collier Cameron 2002).
In particular, the observations of stars in open clusters have shown that
the Skumanich law overestimates the spin down rate to an age
of about $10^8$ yr and, thus, does not explain the presence of fast rotators
in the Pleiades (Stauffer 1987; Andronov et.al 2003). As a possible 
resolution to this problem, MacGregor \& Brenner (1991) suggested that magnetic braking 
is reduced at high rotation rates due to the possible saturation of a stellar dynamo 
resulting in an angular momentum loss rate proportional to $\Omega$. Here, the 
strength of magnetic field at the stellar surface becomes independent of the angular
velocity in the convective zone at high rotation rates.  
The saturation regime is normally considered to occur at rotational rates of about
$10 \Omega_{\odot}$, where $\Omega_{\odot}$ is the angular velocity of the Sun.
While the magnetic field strength in active regions of the Sun is a few orders
of magnitude higher than the average, the fundamental reason for saturation of the 
magnetic field at the rotational velocity which is only one order of magnitude higher
than in the Sun is not understood.  Another possible mechanism presumes that the  
depression of the rate of angular momentum loss may be related to the appearance of 
an increasingly complex field topology as the rotation rate increases leading to a 
reduction in the fraction of open field lines (e.g., Taam \& Spruit 1989).

Observations of CVs also do not provide direct support for an angular momentum loss
rate proportional to a linear function of the angular velocity (see Andronov et al.
2003).  The empirically determined mass transfer rates in CVs have been
estimated to follow the relation

\begin{equation}
\log \dot M = 3.3 \log P_{\rm h} - 11.2 \pm 1 \ ,
\end{equation}

\noindent where $\dot M$ is the mass transfer rate in $M_{\odot}$ yr$^{-1}$ and $P_{\rm h}$
is the binary orbital period in hours respectively (Patterson 1984). Although the inferred mass
transfer rates are somewhat uncertain, they lie
systematically higher than one predicts from the application of the angular momentum
loss rate associated with saturated magnetic braking.

It is therefore clear that there is further need to investigate MB for fast rotators.
Accordingly, we attempt to provide additional insight into the angular momentum loss 
rate associated with the MB process in the context of the  
idealized MS87 model in conjunction with recent observations for the X-ray emission of 
rotating dwarfs (Pizzolato et al. 2003). With these observational constraints, we present
a modified angular momentum loss description and apply it to evolutionary calculations in 
a CV-type close binary system. In \S 2, the model is described and its application to
the evolution of CVs is presented.  We summarize our results and discuss the possible
implications of the modified angular momentum loss rate in the final section.

\section{MODEL of MB}

We consider MB within the framework of the two-component model described in MS87.  In contrast 
to the earlier model described by Weber \& Davis (1967), a dipole magnetic field, in addition 
to a radial field, and centrifugal acceleration effects on the wind are included. The 
strong field regions of this dipole field are envisioned to confine hot plasma in a
corotating dead zone (Mestel 1968).  Although the X-ray emitting corona is largely produced
by matter confined in loops (Vaiana, Krieger, \& Timothy 1973) which fluctuate greatly in the 
Sun's magnetosphere (Sheeley \& Golub 1979; Shimizu \& Tsuneta 1977),
we shall assume that the average behavior in this time dependent 
structure can in the lowest order be approximated by a steady hydrostatic dead zone model.
Partial support for such a simple two-component coronal structure is 
provided by the observations taken during solar eclipse using the {\it Skylab} telescope
and {\it Yohkoh} X-ray satellite.  Since the matter in the dead zone is trapped within 
the magnetosphere and is not lost in the wind, the efficiency of MB in this idealized 
model is reduced in comparison to other idealized theoretical models without such regions.  
The dipole field in the model is assumed to be closed up to a  
cusp point that defines the radius of the dead zone, $R_{\rm d}$. Beyond this point the magnetic
field is assumed to be essentially radial. Recently, support for the large scale 
description of such a picture has been 
provided by the X-ray observations of the Sun obtained from the {\it Ulysses} mission.
In particular, Li (1999) showed that the assumption of homogeneity in latitude of the wind flux 
and the radial magnetic field at large distances, a basic feature inherent in the model, is 
basically correct.

\subsection{Constraints on the coronal density from $L_{\rm x}$}

It was noted in MS87 that the two-component magnetic field model should be consistent  
with the X-ray observations of single stars.  Recent results by Pizzolato et al. (2003) 
confirm earlier results by Pallavicini et al. (1981), showing that the X-ray luminosity, 
$L_{\rm x}$, is proportional to $\Omega^2$ for slowly rotating stars.  Since the X-ray emission 
is related to the coronal gas density in the dead zone, $\rho_{0,\rm d}$, in this model, 
$\rho_{0,\rm d}$ should be a function of $\Omega$. In the simplest model consistent with
the X-ray observations, MS87 considered the possibility that $\rho_{0,\rm d}$ is linear in
$\Omega$.  This choice was based on the assumption that the observed variation of coronal 
emission is determined primarily by the variation in density, that is,  
$L_{\rm x} \propto \rho_{0,\rm d}^2$ and , hence, $L_{\rm x} \propto \Omega^2$,
neglecting the weak dependence on coronal temperature. 
Such a picture is an oversimplification since the variation in X-ray 
luminosity not only reflects variations 
in the  coronal gas density, but also the volume of the X-ray emitting zone. 

Further compounding the validity of the linear relation between the coronal gas density 
and rotational angular velocity are the recent observations of X-ray emission 
from late-type dwarfs, which have shown that the relationship between the X-ray luminosity 
and rotation rate changes form at high rates (Pizzolato et al. 2003).  In particular,
the X-ray luminosity is found to be nearly independent of $\Omega$ for $\Omega \gta 
\Omega_{\rm x} \sim 2 - 12 \Omega_{\odot}$ in the mass range of $0.22 - 1.29 \msun$.
The qualitative change from $L_{\rm x} \propto \Omega^2$ to $L_{\rm x}$ independent 
of $\Omega$ (for $\Omega > \Omega_{\rm x}$) provides an important constraint on the MB 
model.  In the following we make use of this observed trend to constrain the variation
of $\rho_{0,\rm d}$ with $\Omega$ in the MS87 model.

In order to estimate the X-ray luminosity from the MS87 model, we assume that 
the emitting region can be roughly identified with the dead zone in some spatially 
and time averaged 
sense.  The radius of the dead zone is taken from eq.~$(8)_{\rm MS}$ in MS87, where 
it is assumed that $\rho_{0,\rm d} \propto \Omega^p$ and $B_0 \propto \Omega$. In 
addition, we assume that the parameter $\zeta_d = B_0^2/8 \pi \rho_{0,\rm d} 
a_{\rm d}^2$ is equal to 60 where $a_{\rm d}$ is the sound speed in the dead zone (see MS87). 
For a solar type star of mass, $M$, equal to $1 \msun$ the parameter $p$ was varied such that 
the $L_{\rm x}$ varied with $\Omega$ in approximate accordance to the observed relationship, 
with a normalization chosen to correspond to the solar X-ray luminosity at $\Omega_{\odot}$.
The functional form of $L_{\rm x}$ with respect to $\Omega$ was found to be very sensitive
to $p$ for $\Omega \lesssim 10\, \Omega_\odot$, varying as $\Omega^{1.2}$ and $\Omega^{2.8}$ 
%for the coronal gas density in the dead zone chosen to be independent of the 
%angular velocity and a linear function of the angular velocity respectively. 
for $p$ equal to 0 and 1 respectively. 
The relation between $L_{\rm x}$ and $\Omega$ is illustrated in the upper 
panel of Fig. 1, where the best fit for $L_{\rm x} \propto \Omega^2$
corresponds to $p\approx0.6$.

\begin{figure}
\plotone{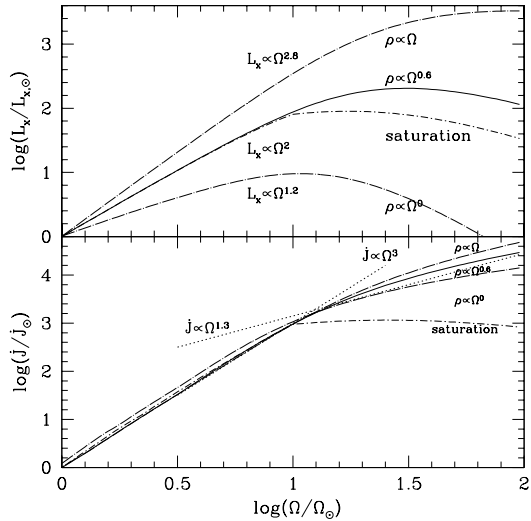}
\caption{The variation of the X-ray luminosity and angular momentum loss rate for a $1 \msun$ 
star as a function of angular velocity. Upper panel: The variation of $L_{\rm x}$, as a function of
$\Omega$.  $L_{\rm x,\odot}$ corresponds to $L_{\rm x}$
at $\Omega=\Omega_\odot$. The thick solid line represents the solution for which  
$\rho\propto \Omega^{0.6}$.
Thick dash-dotted line represent the solution for the same density variation but
with the saturation of the magnetic field at $\Omega \ge 10\,\Omega_\odot$. The 
upper and lower thin dash-dotted lines show $L_{\rm x}$ for  $\rho\propto \Omega^0$
and  $\rho\propto \Omega$ respectively. 
Lower panel: The variation of $\dot J$. as a function of $\Omega$
$\dot J_\odot$ corresponds to the rate of the 
angular momentum loss for the star rotating with  $\Omega=\Omega_\odot$. 
Lines as described for the upper panel. 
}
\end{figure}

Fig. 1 also reveals that $L_{\rm x}$ is
relatively insensitive to angular velocity at high angular velocities, especially for 
$p \gta 0.6$. 
%$\Omega \gta 10 \Omega_{\odot}$.  
This tendency toward saturation occurs 
at high rotation rates since the centrifugal effects become increasingly important, resulting
in the shrinkage of the dead zone (MS87).  The decrease in volume tends to offset the increase in 
$\rho_{0,\rm d}$ as $\Omega$ is increased resulting in a nearly constant value for $L_{\rm x}$. 
For this model, $L_{\rm x}$ actually decreases for sufficiently large $\Omega$. 
At larger values of $p$, $L_{\rm x}$ increases monotonically since the density 
dependence more than compensates for the 
reduction in emitting volume.  On the other hand, 
the volume effect is more important for smaller 
%The volume effect is more important for smaller 
values of $p$, leading to a significant reduction in $L_{\rm x}$ as $\Omega$ is 
increased. 

The ratio of $L_{\rm x}^{\rm sat}$ in the saturated regime, 
taken to correspond to the maximum value of $L_{\rm x}$, to $L_{\rm x,\odot}$
at $\Omega=\Omega_\odot$ for $p= 0.6$ is in approximate accord to that observed.  Specifically, 
$\Delta = \log L_{\rm x}^{\rm sat}/L_{\rm x,\odot} = 2.3$
from the numerical results for $p=0.6$ and  $\Delta  = 2.4$ from the  
observations (see  Fig.~5 in Pizzolato et al. 2003). We point out that the qualitative 
behavior of $L_{\rm x}$ as a function of $\Omega$ is insensitive to the adopted choice
of $\zeta_d$. Quantitatively, the detailed results only change slightly.  For example, for 
$\zeta_d = 4$ (see MS87), $p \sim 0.4$ and $\Delta = 2.2$. 

For comparison, we have also displayed the solution with a saturated magnetic field
for $\Omega \ge 10\,\Omega_\odot$ in Fig. 1.  Although $L_{\rm x}$ is seen to turnover 
as well, we note that it occurs at a level significantly less than observed with $\Delta  
= 2$. We cannot exclude the possibility that saturation of the magnetic field takes 
place based on the observed relation $L_{\rm x}(\Omega)$ since it may take place at 
rotation rates higher than $10 \Omega_\odot$ making it difficult to distinguish 
the influence of the saturation of the magnetic field from the influence
of the variation of the dead zone with angular velocity. However, the X-ray observations 
of solar type stars alone do not require the additional assumption of a saturated 
magnetic field at high rotation rates.

\subsection{MB rate}

The transport of the angular momentum by the wind and by Maxwell stresses 
is equivalent to that found by assuming corotation maintained out
to the Alfven surface (Mestel 1984). 
The angular momentum loss rate can be written as

\begin{equation}
-\dot J = 4 \pi\Omega \int_0^{\pi/2} (\rho_{\rm A} v_{\rm A} R_{\rm A}^2) 
(R_{\rm A} \sin \theta)^2 \sin \theta d \theta,
\end{equation}

\noindent where $\theta$ is the polar angle, and $\rho_{\rm A}$, $v_{\rm A}$, and 
$R_{\rm A}$ are the density, velocity, and radius at the Alfven surface respectively.
By continuity, $\rho_{\rm A} v_{\rm A}/\rho_{\rm d} v_{\rm d} = 
B_{\rm A}/B_{\rm d} = (R_{\rm d}/ R_{\rm A})^2$ assuming a radial field 
beyond the dead zone region. Here the radius od the dead zone is 
found from equ.~$(8)_{\rm MS}$ of MS87.
Then

\begin{eqnarray}
-\dot J &=& 4 \pi\Omega_\odot R_\odot^4 \rho_{\rm d,\odot} v_{\rm d, \odot}
\left ( {\frac {v_{\rm d}} {v_{\rm d, \odot}} } \right )
\left ( {\frac {R} {R_\odot} } \right )^4
\left ( {\frac {\Omega} {\Omega_\odot} } \right )^{p+1} \nonumber \\ 
 & &\int_0^{\pi/2} 
\left ( {\frac {R_{\rm A}} {R} } \right )^2
\left ( {\frac {R_{\rm d}} {R} } \right )^2
\sin^3 \theta d \theta \nonumber \\ 
&=& C_j \left ( {\frac {v_{\rm d}} {v_{\rm d,\odot}} } \right )
\left ( {\frac {R} {R_\odot} } \right )^4
F \left ( M, {\frac {\Omega} {\Omega_\odot} } \right ).
\label{eq:mbrate}
\end{eqnarray}

\noindent where $R$ is the stellar radius and $C_j$ incorporates 
all physical constants.\footnote{$F(M, \Omega/\Omega_\odot)$ also
depends on the solar sonic velocity and the coronal density, 
but we treat them as constants in order to scale the equation to the observed 
value of the solar Alfven radius and the dead zone. 
We are aware that $R_{\rm A,\odot}$ and $R_{\rm d,\odot}$ are not known 
precisely, and in particular are different in MS87 and Li(1999).
Since we do not attempt to derive the precise value for the current 
solar $\dot J$ directly from observations as in Li (1999), 
we adopted these values as in MS87 and assume that uncertainties 
in the measured solar values are absorbed in the calibration of $\dot J$.}
The Alfven radius is found from equ.~$(14)_{\rm MS}$ of MS87.

In the lower panel of Fig.~1, the angular momentum loss rate relative to the Sun, 
$\log(\dot J / \dot J_\odot)$, is shown as a function of normalized angular velocity, 
$\Omega / \Omega_{\odot}$, for three different values of $p$ corresponding to $p=0,
\,0.6\ {\rm and}\ 1$.  For reference, we also illustrate the variation of the angular 
momentum loss rate for the case of a saturated magnetic field (for $\Omega > \Omega_x$) 
with $p=0.6$. For slow rotation, the MB is consistent with the Skumanich law and is
nearly independent of the dependence of the coronal density on angular velocity 
(in contrast to the variation of $L_{\rm x}$ with $\Omega$).
However for fast rotation ($\Omega \gtrsim 10\,\Omega_\odot $), $\dot J \propto 
\Omega ^{1.3}$ for $p=0.6$, the value inferred from the constraints based on the 
X-ray observations (see above). We note that the asymptotic form of the angular 
momentum loss rate on the angular velocity, in contrast to the X-ray luminosity, is 
insensitive to the value of $p$ if $p$ lies between 0.6 and 1.  On the other hand, 
if the magnetic field is saturated, the angular 
momentum loss rate is approximately constant at high rotation rates.  Hence, the 
loss rate from a centrifugally driven wind 
%in the latter case 
differs from the linear functional form suggested by MacGregor \& 
Brenner (1991) for saturation of the stellar dynamo based on a thermally driven 
stellar wind model. 

Based on these results, the solution for  the rate of angular momentum loss due to MB 
can be presented in the parameterized form

\begin{equation}
- \dot J = K_j \left ( {\frac {R} {R_\odot} } \right )^4
 \left ( {\frac {T_{\rm d}} {T_{\rm d,\odot}} } \right )^{1/2}
\left\{  \begin{array}{c c}
(\Omega/\Omega_\odot)^3 \, \, \, \, \, \, \, \ ,{\rm for\ } \Omega \le \Omega_{\rm x} \\ 
 \Omega^{1.3} \Omega_{\rm x}^{1.7}/\Omega_\odot^3,{\rm for\ } \Omega > \Omega_{\rm x} 
\end{array} \right. 
\end{equation}

\noindent Here, the value of $K_j$ depends on the mass of the star,  and the temperature 
and density in the 
corona which can be derived directly from the observational data (Li 1999). 
This parametrized form is presented only to provide a more transparent form of the 
angular momentum loss rate for 
comparison between the magnetic braking law based on equ.~{\ref{eq:mbrate}}
and the power law representations describing the Skumanich law (e.g., as in Verbunt \& Zwaan 1981)
or the saturated braking law (Andronov et al. 2003). In detailed calculations, 
the rate of angular momentum loss due to MB should be found using equ.~{\ref{eq:mbrate}}
together with highly nonlinear equ.~$(8)_{\rm MS}$ and equ.~$(14)_{\rm MS}$ of MS87.

In this study we adopt the approach of calibrating $K_j$ using evolutionary calculations
to reproduce the known solar rotation period at the age of the Sun (see Kawaler 1988).
For $\Omega_\odot=3 \times 10^{-6}$ we find
$K_{j} = 6\times 10^{30}$ [dyn cm],
which is about three times the value obtained in Li (1999) and about the same as in
Pylyser \& Savonije (1988) (their value includes a variable factor $0.73 <f<1.78$ ).
The corresponding value for the constant in the equ.~(4) is
$C_j = 2.1\times 10^{27}$ [dyn cm]. 
\footnote{Our code does not incorporate the angular velocity in the 
equation of hydrostatic equilibrium and therefore a more accurate
calibration can be carried out with an evolutionary code including the effects of rotation.
However, since the value of $K_j$ mainly depends on the evolution of a relatively slowly 
rotating star, where $\Omega$ does not contribute significantly
to hydrostatic equilibrium, our results should be adequate.}

Our description of magnetic braking has been applied to a study of the rotational evolution 
of single stars for comparison to the observed  
spin-down of stars in young clusters. Although stellar rotation is not 
included in the stellar structure, we calculated two sets of stellar models of mass 
1, 0.8 and 0.6 $\msun$ with   
the chemical composition as in Pleiades and Hyades to obtain an indication of the effect 
of our modified braking rate.  The models were evolved for a time corresponding to 
$7 \times 10^7$ yr for the Pleiades and $6 \times 10^8$ yr for the Hyades clusters.
The initial surface rotational velocity was taken as the break-up velocity.
The numerical results as presented in Table 1 show that the level of agreement between the 
observations and the angular momentum loss rate presented in this study and with the loss 
rate based on a saturated model is comparable. The agreement is good for the Pleiades  
cluster and is slightly better for stars of 
mass 1 and 0.8 $\msun$ in comparison to the lower mass model for the Hyades cluster.  Here,  
the observational data for $v \sin i$ and data for the spin-down based on the saturated 
magnetic braking law
and the Skumanich law have been taken from Andronov et al. (2003).

\begin{deluxetable}{l l l l l}
\tabletypesize{\scriptsize}
\tablecaption{
The rotational velocities obtained from the theoretical models and from the observations
for the Pleiades and Hyades clusters, in km s$^{-1}$. Here, ``observed'' corresponds to the
maximum observed velocity of stars in the Pleiades or the Hyades, ``saturated'' represents
the results of calculations based upon the saturated
magnetic braking law, and ``standard'' represents the maximum rotational velocity that can be
obtained with the standard prescription for the magnetic braking (all the data is taken 
from Andronov et al. 2003); ``this work'' represents results of the spin-down using 
equ.{\ref{eq:mbrate}}.
}
\tablehead{\colhead{cluster}     & \colhead{velocity}      & \colhead{$1.0M_\odot$} &
	   \colhead{$0.8M_\odot$} &\colhead{$0.6M_\odot$}}
\startdata
 Pleiades & observed   & 60 & 50 & 100\\
          & this work  & 70 & 95 & 140 \\
          & saturation & 100 & 105 &  100 \\
          & standard   & 15 & 20 & 25 \\
 Hyades   & observed   & 5 & 4 & 9\\
          & this work  & 6 & 6 & 6 \\
          & saturation & 6 & 7 &  9 \\
          & standard & 6 & 6 & 8 \\
\enddata
\label{table}
\end{deluxetable}

\subsection{Binary evolution}

To determine the consequences of our modified angular momentum loss rate prescription on 
the evolution of a binary system, we calculated the mass transfer phases of two binary 
model sequences.  The model stars are chosen to be of a solar metallicity.  The first 
sequence consisted of 1 $M_{\odot}$ evolved donor and the second a 0.8 $M_{\odot}$
unevolved donor, both are on the MS and, with a 0.6 $M_{\odot}$ white dwarf companion.  Such types of systems 
can be considered to be representative of CVs.  For the evolved donor, the central 
hydrogen abundance was $X=0.35$.  The radii of the evolved and unevolved model stars 
were such that they filled their Roche lobes at orbital periods of $P=7.5$ h and
5.2 h respectively.  To follow the evolution, we used the code described in Ivanova 
et al. (2003), supplemented with angular momentum loss rates associated with MB based on 
equ. (3). The evolution of the system is assumed to be non conservative with nearly all the 
mass lost from the donor{\footnote{For details on the treatment in our code for the WD evolution and its 
ability to accumulate mass see Ivanova \& Taam in preparation (see also Li \& van den 
Heuvel 1997).}
ejected as a consequence of nova explosions or as a result 
of an optically thick wind from the white dwarf surface (Hachisu, Kato, \& Nomoto 1999). 
For comparison, we also performed mass transfer sequences with MB according 
%%Natasha - added a specific reference here - This is what Pylyser and Savonije used 
to Verbunt \& Zwaan (1981) based on
the Skumanich law (taken from Pylyser \& Savonije, 1988).

The results for the mass transfer rates are shown as a function of orbital period in 
Fig. 2.  The average mass transfer rate for both donors is $\sim 5 \times 
10^{-10} \mpy$. 
For comparison, we also display the mass transfer rates corresponding to those
observationally estimated according to an empirical relation obtained by Patterson (1984).  
It can be seen that the mass transfer rates promoted by the Skumanich law are generally
higher than the empirical values, and the rates produced by the modified MB rate are in 
approximate agreement. The evolutionary timescale of the systems is increased by about
a factor of 5 for the both sequences with the 
modified MB rate in comparison to the prescription based on the Skumanich law.  We 
note that this is significantly less than the factor of 100 found by Andronov et al. (2003) 
in their case for MB rate based on $\dot J \propto \Omega$.

\begin{figure}
\plotone{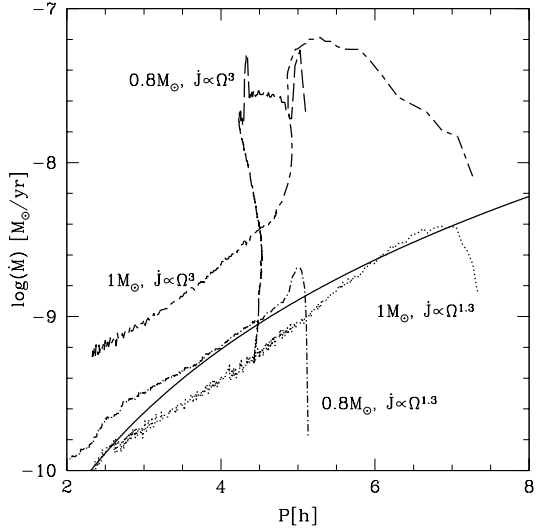}
\caption{The variation of mass transfer rate (in $\mpy$) as a function of orbital period 
(in hours) for two model sequences.  The evolution of a 1 $M_\odot$ evolved donor
and  0.6 $M_\odot$ white dwarf is illustrated by a dotted line for the angular momentum 
prescription presented in this paper and as a dash dashed line for the angular momentum prescription 
assuming that the Skumanich law applies to high angular velocities.  The dot dashed and dashed line 
illustrate the evolution of a $0.8 \msun$ unevolved donor with a $0.6 \msun$ white dwarf with 
the prescription adopted here and the Skumanich law respectively.  For reference, the solid 
line represents the empirically inferred mass transfer rates obtained by Patterson (1984).}

\end{figure}

\section{SUMMARY}

The X-ray luminosity of late type stars and the angular momentum loss rate associated 
with magnetic braking of these 
stars has been examined in terms of an idealized two-component coronal 
model in the light of recent observational studies relating stellar activity to stellar 
rotation.  By assuming that the X-ray luminosity can be used as a proxy for the
coronal gas density in the magnetosphere we have, as a first step, 
constrained the two-component model
to reproduce the observed variation of X-ray luminosity with stellar rotation ($L_{\rm x}
\propto \Omega^2$) for slowly rotating late type dwarfs.  The application of this model to
rapidly rotating dwarfs provides a qualitative explanation for the observed saturation
at high rotation rates where the decrease in volume of the X-ray emitting coronal region
nearly compensates for the increase in coronal gas density with increasing rotational 
velocity. 

The angular momentum loss rate resulting from such a model has been analyzed and a  
parameterized form for the rate has been presented.  In particular, it has the form
$\dot J \propto \Omega^3$ for slow rotators to reproduce the Skumanich law and $\dot J 
\propto \Omega^{1.3}$ for fast rotators ($\Omega > 10 \Omega_{\odot}$).  The angular  
momentum loss rate is more sensitive to $\Omega$ in the high rotation rate regime than 
the form, $\dot J \propto \Omega$, adopted in recent studies by
Andronov et al. (2003). 
In addition, the reduction of the angular momentum loss rate from that based on the  
Skumanich relation for slowly rotating stars, required by the observation of rapidly 
rotating stars in clusters, can be accommodated without necessarily invoking the assumption 
that the magnetic field saturates at high rotation rates.  We note, however, that the 
observations of rotational velocites of stars less massive than about $0.4 \msun$ 
(see Andronov et al. 2003 and references therein) suggest
that the angular momentum loss rate is further weakened. 

The modified form of the angular momentum loss rates can be applied to both 
LMXBs and CVs, and we have carried out binary evolutionary calculations for
model systems representative of CVs as an illustration.  It has been shown that
the mass transfer rates are approximately consistent with those observed for CVs 
provided that the observed rates are accurate and reflect the long term secular values.   The 
calculated rates for unevolved and evolved main sequence-like stars are found to be about an 
order of magnitude lower than rates obtained using angular
momentum loss rates based on the application of the prescription for slowly
rotating stars to the rapidly rotating stars in CVs and are higher by a factor
of 10 than the rates obtained using rates obtained using a saturated braking law
(Andronov et al. 2003).  
As a consequence the time scale
for both the pre CV's to evolve into the CV phase and the evolution during the CV phase 
are increased in comparison to previous detailed binary evolutionary studies, but a factor 
of 10 shorter than obtained using a saturated law.  Hence, the possibility that finite
age effects are important in understanding the distribution of such systems (e.g.,
King \& Schenker 2002; Andronov et al. 2003) is made less likely. 

%An additional consequence of the reduced level of angular momentum loss rate for rapidly 
%rotating stars is a decrease in the binary orbital period delineating those systems 
%which evolve to shorter periods from those 
%which evolve to longer orbital periods.  That is, angular momentum losses dominate 
%the evolution of the binary for periods less than the bifurcation period whereas 
%nuclear evolution dominates the evolution of the system at longer orbital periods
%(see Pylyser \& 
%Savonije 1989).  Upon comparison of the angular momentum loss timescale with the nuclear
%evolutionary timescale, we estimate that the bifurcation period for a $1 \msun$ 
%compact object is about 1.6 days.  

Although the mass transfer rates are significantly higher than that obtained 
by Andronov et al. (2003), the rates are a factor of 2 to 3 times lower than the 
rates ($\dot M > 1.5  - 2\times 10^{-9} \msun$ yr$^{-1}$)
required to force an unevolved donor sufficiently out of thermal equilibrium to explain
the period gap width in the disrupted magnetic braking model (see McDermott \& Taam
1989; Kolb 2002). As noted previously, the upper boundary of the period 
gap at about 3 hours approximately corresponds to a main sequence star of $\sim 0.4 
\msun$. However, there is no observational evidence indicating a discontinuous change in the 
rate of spin down of single stars, although, for less massive stars, the rotational 
velocity data of clusters suggests a further weakening of the angular momentum loss
rate (see Sills, Pinsonneault, \& Terndrup 2000).  This suggests that 
other mechanisms should be sought to prevent 
systems from entering into the period gap.  A possible alternative involves a  
period bounce first suggested by Eggleton (1983) involving additional angular
momentum losses, perhaps, beyond that considered in the present study (see e.g., Taam, Sandquist, 
\& Dubus 2003).  

An additional consequence of the reduced angular momentum loss rate for rapidly 
rotating stars will be a revision of the binary orbital period delineating those systems
which evolve to shorter periods (angular momentum loss dominated evolution) from those 
which evolve to longer orbital periods (nuclear evolution dominated).
This period not only depends on the rate of angular momentum loss, but is also a 
function of the mass of the binary components and the degree to which mass is lost 
from the system.  For the case of conservative mass transfer, 
Pylyser \& Savonije (1988, 1989) estimate a bifurcation period of about 
12 hours for systems with an accreting compact companion.  With a reduced effectiveness 
of magnetic braking, it is expected that the transition will shift to shorter orbital periods. 

Future phenomenological studies of the X-ray emission from late type stars and the 
angular momentum losses associated with magnetic braking in these stars are 
desirable to determine the explicit mass dependence on the input parameters (magnetic 
field, density and temperature of the magnetically confined region) in the 
simplified magnetically coupled stellar wind models.  The constraints imposed by 
both the X-ray studies and rotational velocity studies may provide
insight into the further reduction of the angular momentum loss rates for stars less 
massive than $0.4 \msun$.  The incorporation of these ideas in binary evolutionary 
sequences will be necessary 
in order to make meaningful comparisons between the 
observed and theoretically predicted CV period distributions in future population 
synthesis studies.

\acknowledgments
We would like to thank Dr. H.~Spruit for useful discussions.
This work is partially supported by the NSF through grant AST-0200876.

%\vfil\eject

%\vfil\eject

\end{document}